\documentstyle[prb,aps,epsf]{revtex} \def\narrowtext{} \tighten \twocolumn
\input epsf.sty

\begin{document}
\draft

\title{Magnetic Collective Mode Dispersion in High Temperature Superconductors}
\author{M. R. Norman}
\address{Materials Science Division, Argonne National Laboratory, 
Argonne, IL  60439}

\address{%
\begin{minipage}[t]{6.0in}
\begin{abstract}
Recent neutron scattering experiments in the superconducting state of
YBCO have been interpreted in terms
of a magnetic collective mode whose dispersion relative to the commensurate
wavevector has a curvature opposite in sign to a conventional magnon dispersion.
The purpose of this article is to demonstrate that simple linear response
calculations are in support of a collective mode interpretation, and to explain
why the dispersion has the curvature it does.
\typeout{polish abstract}
\end{abstract}
\end{minipage}}

\maketitle
\narrowtext

Inelastic neutron scattering, which measures the momentum and energy
dependence of the dynamic spin susceptibility, has emerged as a powerful probe
of high temperature cuprate superconductors\cite{FONG2}.  One of the most
striking features of the data is the appearance below the superconducting
transition temperature of a commensurate resonance in the case of the
bilayer cuprates, YBCO and Bi2212\cite{ROSSAT,MOOK,FONG,KEIMER}.  In addition,
an incommensurate magnetic response has been seen for energies below the
resonance energy in both cases\cite{MOOK2,MOOK4,ARAI}.  Recently, it has
been argued that these two structures are part of the same collective mode
dispersion\cite{BOURGES}.  This interpretation is unusual in that the curvature
of the purported collective mode dispersion relative to the commensurate
wavevector is opposite in sign to that of a conventional magnon dispersion.

One of the motivations of this interpretation is the similarity of the
measured dispersion to theoretical results based on linear response theory
(RPA) \cite{BLEE,AC,FLORA,KAO,NORM00}.  In these works, though, it was
unclear whether the incommensurate response below resonance was actually
due to a pole in the response function as would be expected for a collective
mode interpretation.  The purpose of this article is to investigate this
issue in detail.

The methodology is standard, and details can be found in
Ref.~\onlinecite{NORM00}.  An effective quasiparticle dispersion is used to
calculate the bare bubble in the superconducting state, $\chi_0$. The RPA 
susceptibility is then obtained as $\chi_0/(1-J\chi_0)$.  The most 
crucial input is the form of the dispersion.  In Ref.~\onlinecite{NORM00}, two
dispersions were analyzed, one based on angle resolved photoemission (ARPES)
data in the normal state 
(tb1), another on ARPES data in the superconducting state where the Fermi 
surface was flattened around the node to enhance the effects of
incommensurability (tb2).  
In addition, in this paper, a slight modification of this second 
dispersion was made to further enhance the incommensurability (tb3).  All three 
dispersions are listed in Table I for completeness sake.  For the 
superconducting gap, the standard $\cos(k_x)-\cos(k_y)$ form was assumed, 
with a maximum value, $\Delta_{max}$, taken from recent tunneling measurements
in YBCO\cite{WEI}.  $J$ was adjusted to 
obtain a resonance at 41 meV, the resulting $J$ being similar in value to 
estimates based on neutron scattering in underdoped samples \cite{HAYDEN}.

To begin, it is helpful to discuss the frequency dependence of the
bare bubble, $\chi_0$.  In Figure 1, $Re\chi_0$ and $Im\chi_0$ are shown at
a typical incommensurate wavevector for the three dispersions of Table 1.
In $Im\chi_0$, there are two steps, a low energy one at threshold, and another
at slightly higher energy.  The two energies correspond to the spin
gap energies $\omega_g(\pm q)=\min_k(E_k+E_{k \pm q})$, where $k$ is confined
to the first quadrant of the Brillouin zone and $E_k$ are the quasiparticle
energies in the superconducting state.  The steps are a consequence
of the coherence factors associated with a d-wave order parameter.  For the
commensurate wavevector, $Q=(\pi,\pi)$, these two energies are degenerate,
so there is
only one step.  By the Kramers-Kronig relations, these steps lead to
logarithmic divergences in $Re\chi_0$ which become simple peaks when
damping is taken into account.  (All plotted results are obtained by
replacing $\omega$ by $\omega+i\Gamma$ in the energy denominators when
calculating $\chi_0$).

In Figure 2, $Re\chi_0$ and $Im\chi_0$ are shown as a function of $q$ at an
energy of 35 meV 
for the three dispersions of Table 1.  In the first case (tb1), one finds 
the typical result that $Re\chi_0$ is quadratic relative to the commensurate
wavevector $Q$ with a negative curvature (i.e., a maximum at $Q$).
Since $Re\chi_0$ increases with $\omega$ (up to the spin gap energy
$\omega_g(Q)$), one expects that the RPA pole condition, $1-JRe\chi_0=0$,
will be satisfied first at $Q$, then at higher energies at
wavevectors increasingly displaced from $Q$.  This would then give a
conventional magnon like dispersion.  The actual situation is more
complicated, though.  Note in Figure 2a that there is a second maximum
in $Re\chi_0$
at an incommensurate wavevector, associated with $\omega_g(q)$ as discussed
in the context of Figure 1.  As the energy increases, this second
maximum becomes the global maximum.

In Figure 3a, the $q$ vector where the RPA $Im\chi$ is maximal at a given
$\omega$
($\omega(q)$) is plotted for this dispersion.  In agreement with experiment,
one finds
incommensurability below the resonance energy, but the RPA pole condition is
not satisfied below resonance.  Instead the incommensurability simply
tracks $\omega_g(q)$.  As a consequence, the intensity plummets below
resonance, as can be seen in Figure 4a.  Above resonance, the RPA pole
condition is satisfied, but as can be seen in Figure 3a, the dispersion
crosses into the continuum (i.e., beyond $\omega_g(q)$) just above resonance,
so the mode becomes damped, as obvious from the intensity plot
of Figure 4a.

The above behavior is not what is observed, in that the experimental
intensity decays more slowly below resonance.  This is because the
incommensurability effects associated with dispersion tb1 are too weak.
This can be contrasted with the dispersion analyzed by Brinkmann and
Lee \cite{BLEE}.  In that case, the constant energy contours of $E_k$ are
quite flat for energies below $\Delta_{max}$, leading to an enhancement in the
incommensurate effects.  This
can be seen in Figures 2b and 2c, where $Re\chi_0$ is plotted for the other
two dispersions considered here (tb2 and tb3).  In the first case (tb2),
the dispersion used in earlier work \cite{NORM00}, one finds a rather flat
behavior around $Q$ due to the flat quasiparticle dispersion around
$k=(\pi,0)$.  In addition, a global maximum now occurs at an incommensurate
wavevector due to the flattening of the Fermi surface around the d-wave nodal
$(\pi,\pi)$ directions relative to dispersion tb1.

The resulting dispersion, $\omega(q)$, of $Im\chi$ from tb2 is
plotted in Figure 3b.  This is in better agreement with experimental
results in YBCO \cite{ARAI,BOURGES}.  In addition, it can be seen from
this plot that the RPA pole condition is now satisfied below the resonance
energy for a few meV until the spin gap energy, $\omega_g(q)$, is
encountered.  For energies below this, the mode is strongly diminished
in intensity as it traces out the edge of the continuum.  This result is
consistent with earlier work \cite{FLORA}.

Although Figure 3b is in good agreement with experiment as far as
$\omega(q)$ is concerned, it is still deficient in that the intensity of
$Im\chi$ below resonance still drops off faster than experiment, as plotted in
Figure 4b and also noted earlier \cite{NORM00}.  To analyze this further,
dispersion tb2 was modified by making the dispersion less flat in one
direction around $(\pi,0)$, and flattening the Fermi surface around the
d-wave node even more.  The resulting $Re\chi_0$ is shown in Figure 2c,
and is similar to that obtained earlier by Brinkmann and Lee \cite{BLEE}.
One now finds a truly quadratic behavior around $Q$ with positive 
curvature, a direct consequence of the stronger incommensurate peaks.
As a consequence, the RPA pole condition is satisfied for an even greater
energy range below resonance (Figure 3c), and moreover, the intensity below
resonance falls off much more slowly (Figure 4c), as observed
experimentally.

The results in this paper concentrated on the $(\pi,q_y)$ direction, 
since this is where the maximum in the incommensurability was observed.
Two dimensional $q$ plots from dispersions tb2 and tb3 are consistent
with experiment \cite{MOOK2} in showing a baseball diamond shaped
incommensurability pattern below resonance with global maxima along
$(\pi,q_y)$ and $(q_x,\pi)$, as reported earlier \cite{NORM00}.  Recently,
an anisotropy was observed in this intensity pattern for partially 
detwinned YBCO samples \cite{MOOK5}.  If this is a true
magnetic anisotropy not related to the phonon background, then within the
RPA context, it would have to be due to the influence of the chain bands
wiping out two of the incommensurate spots, not surprising given the
sensitivity of the incommensurability to the electronic structure.  On
the other hand, this is conjecture at this stage, as a proper calculation
taking into account both plane and chain bands has yet to be performed.
Of course, the most natural explanation of this anisotropy would be due
to stripe formation, but it is unclear whether such a picture can reproduce
the results demonstrated here (in particular, the energy dependence
of the incommensurability).

In conclusion, RPA calculations of the dynamic susceptibility are in
support of a collective mode interpretation of the
magnetic dispersion relation observed in bilayer
cuprate superconductors.  The observed incommensurability has strong 
implications for the effective quasiparticle dispersion, and it would be of 
interest to verify whether the dispersions analyzed here, in particular 
in regards to the flattening of the Fermi surface around the d-wave node, are 
consistent with high resolution angle resolved photoemission data, 
particularly in the case of YBCO.

The author thanks Andrei Chubukov and Oleg Tchernyshyov for discussions.
This work was supported by the U.S. Dept. of Energy, Basic Energy 
Sciences, under Contract No. W-31-109-ENG-38.

\begin{figure}
\epsfxsize=3.4in
\epsfbox{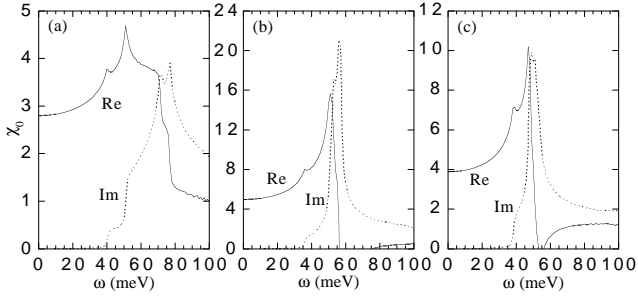}
\vspace{0.5cm}
\label{fig1}
\caption{$Re\chi_0$ and $Im\chi_0$ for $\vec{q}=(1,0.8)\pi$ versus $\omega$ for 
the three dispersions of Table I, with $\Delta_{max}$=29 meV (a and b)
and 25 meV (c),
$\Gamma$=0.5 meV, and $T$=13K.  Units are states/eV/CuO plane,
and so should be multiplied by $2\mu_B^2$ to compare to neutron
susceptibilities.}
\end{figure}

\begin{figure}
\epsfxsize=3.4in
\epsfbox{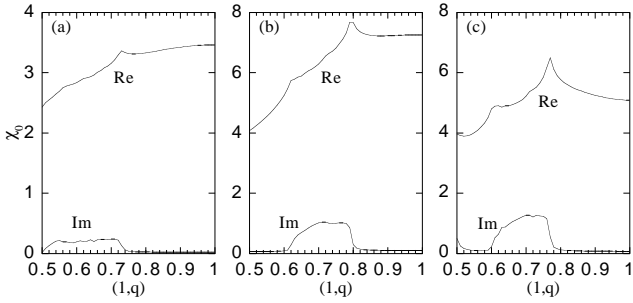}
\vspace{0.5cm}
\label{fig2}
\caption{$Re\chi_0$ and $Im\chi_0$ for $\omega$=35 meV versus 
$\vec{q}=(1,q_y)\pi$.  Same notation as Fig.~1.}
\end{figure}

\begin{figure}
\epsfxsize=3.4in
\epsfbox{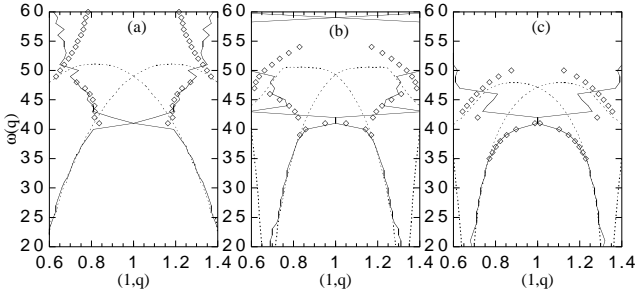}
\vspace{0.5cm}
\label{fig3}
\caption{$\vec{q}=(1,q_y)\pi$ where $Im\chi$ is maximal at a given $\omega$
(solid lines).  The dashed lines are the spin gap energies,
$\omega_g(\pm q)$, defined in the text, and the open diamonds represent
the RPA pole condition, $1-JRe\chi_0(q)=0$.  
Same notation as Fig.~1, with $J$=262 meV (a), 111 meV (b), and 155 meV (c).
Note that $Im\chi_0 \approx 0$ for $\omega$ below the lowest dashed line.}
\end{figure}

\begin{figure}
\epsfxsize=3.4in
\epsfbox{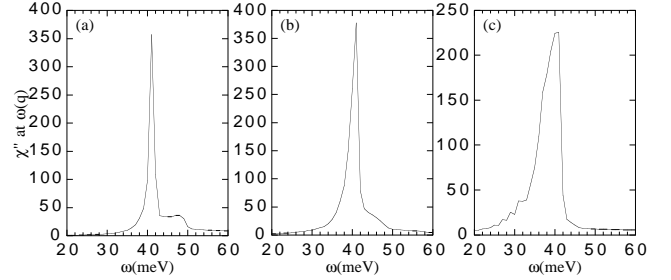}
\vspace{0.5cm}
\label{fig4}
\caption{Value of $Im\chi$ along the $\omega(q)$ dispersion relation from
Fig.~3.}
\end{figure}

\begin{table}
\caption{Tight binding dispersions based on angle resolved photoemission data.
The first three columns list the coefficient, $c_i$, of each term (eV), that is
$\epsilon(\vec k) = \sum c_i \eta_i(\vec k)$, with tb1 and tb2 previously
considered dispersions\protect\cite{NORM00}, and tb3 a modified version 
of tb2 as discussed in the text.  The last column lists the basis
functions (the lattice constant $a$ is set to unity).}
\begin{tabular}{rrrc}
tb1 & tb2 & tb3 & $\eta_i(\vec k)$ \\
\tableline
 0.1305 & 0.0879 & 0.1197 & $1$ \\
-0.5951 &-0.5547 &-0.5881 & $\frac{1}{2} (\cos k_x + \cos k_y)$ \\
 0.1636 & 0.1327 & 0.1461 & $\cos k_x \cos k_y $ \\
-0.0519 & 0.0132 & 0.0095 & $\frac{1}{2} (\cos 2 k_x + \cos 2 k_y)$ \\
-0.1117 &-0.1849 &-0.1298 &  $\frac{1}{2}
 (\cos 2k_x \cos k_y + \cos k_x \cos 2k_y)$ \\
 0.0510 & 0.0265 & 0.0069 & $\cos 2k_x \cos 2k_y $ \\
\end{tabular}
\end{table}

\end{document}